\documentclass[aps,prl,twocolumn,groupedaddress]{revtex4}
\usepackage{graphicx,epstopdf}

\begin{document}


\title{Parametric cooling of a degenerate Fermi gas in an optical trap}


\author{Jiaming Li, Ji Liu, Wen Xu, Leonardo de Melo, and Le Luo}
\email[leluo@iupui.edu]{}
\affiliation{Department of Physics, Indiana University Purdue University Indianapolis, Indianapolis, IN 46202}


\date{\today}

\begin{abstract}
We demonstrate a novel technique for cooling a degenerate Fermi gas in a crossed-beam optical dipole trap, where high-energy atoms
can be selectively removed from the trap by modulating the stiffness of the trapping potential with anharmonic trapping frequencies. We measure the dependence of the cooling effect on the
frequency and amplitude of the parametric modulations. It is found that the large anharmonicity along the axial trapping potential allows to generate a degenerate Fermi gas
with anisotropic energy distribution, in which the cloud energy in the axial direction can be reduced to the ground state value.
\end{abstract}


\maketitle

Evaporative cooling in an optical dipole trap (ODT) has remained a key technique for producing Bose-Einstein condensates and degenerate
Fermi gases for more than a decade~\cite{ChampmanBEC,GranadeAllopt,Grimmreview}. The most common approach for evaporation is to reduce the optical
trapping potential continuously by decreasing the intensity of the trapping beams, so called the ``weakening'' scheme.
The weakening scheme results in a reduction of trapping frequencies inevitably, which not only decreases the collision rate but also limits the maximum
phase space density available in an optical trap. To overcome this drawback, several auxiliary techniques have been implemented to maintain trapping frequencies
during evaporation, including a dimple trap~\cite{Clement09}, moving traps~\cite{Kinoshita05}, time-delay traps~\cite{Arnold11}, and a magnetic field tilting trap~\cite{Hung08}.
These techniques increase the evaporation speed and the final phase space density substantially, but require a more experimental setting.

Alternatively, it is desirable to develop an  ``expelling'' scheme for an ODT, an analogy of the radio-frequency knife for a magnetic trap~\cite{KetterleVanDruten}, where high-energy atoms can be selectively removed from optical traps while keeping the trapping potential intact. Since both the collision rate and the phase space density scale with the cube of the average trapping frequency~\cite{Luo06Cooling},
such an expelling scheme has the potential to improve evaporative cooling in optical traps significantly, which will be essential for experiments with ultracold polar molecules. In those experiments, the coldest sample is close to the Fermi temperature $T_F$ in an ODT, but cooling into deep quantum degeneracy has yet to be realized~\cite{Moses15}. Developing an expelling scheme may pave the way for the final stage cooling in the degenerate regime.

\begin{figure}[tb]
\includegraphics[width=3.3in]{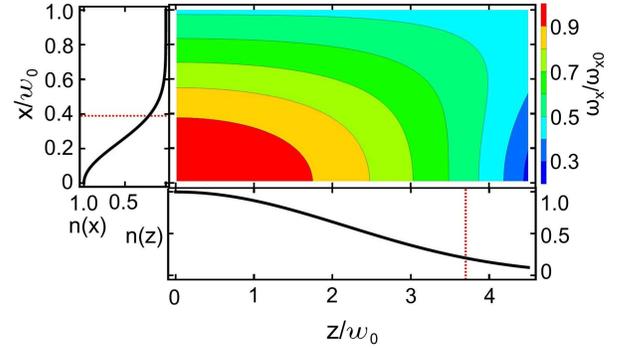}
\caption{The local radial trap frequency of a crossed-beam optical trap. The x-axis trap frequency $\omega_x(x,0,z)$ is plotted in the x-z plane in term of the
harmonic frequency $\omega_{x0}=760$ Hz (the calculated value from the trapping potential). The radial and axial atom densities $n(x)$ and $n(z)$ are plotted in the left and bottom frames for a Fermi gas of
$1.6\times10^5$ atoms per spin state at $T/T_F=0.6$ using 1D Thomas-Fermi distribution~\cite{luothesis}. The red dashed lines show the positions of the Fermi radii in the radial ($\sigma_x$) and axial ($\sigma_z$) directions, where the local trapping frequency drops to $\omega_x(\sigma_x,0,0)=0.89\,\omega_{x0}$ and $\omega_x(0,0,\sigma_z)=0.55\,\omega_{x0}$.}
\label{fig:2DAnH}
\end{figure}

In this letter, we report an ``expelling'' scheme to cool a degenerate Fermi gas by parametric excitation of high-energy atoms
out of an optical trap. Our scheme employs the intrinsic anharmonicity of a crossed-beam ODT, where high-energy atoms
experience smaller trapping frequencies than low-energy atoms. The spatial differential trapping frequencies turn parametric excitation of atomic motion
from a well-established laser-induced heating and loss source~\cite{savard97,gehm98} into a robust cooling mechanism, in which high-energy atoms can be selectively removed from the trap when
the modulation frequency is tuned to resonance with the trapping frequencies of high-energy atoms. Parametric modulation induced cooling has previously been observed for bosonic atoms either in a
magnetic trap~\cite{Kumakura03} or in a standing wave lattice~\cite{Poli02}. However, in both cases, the bosonic atoms were in the thermal states with
phase space densities of $10^{-4}\sim10^{-7}$. When approaching the quantum regime with a phase space density close to one, bosonic atoms tend to
occupy the lowest vibrational states, resulting in a negligible differential trapping frequency between high-energy and low-energy atoms. It
becomes very difficult to parametrically cool a Bose gas at very low temperatures, which has not yet been reported, to the best of our knowledge.
In contrast, fermionic atoms, indebted to the Pauli exclusion principle, occupy a significant fraction of
the vibrational states even at the degenerate temperature, making parametric cooling much more feasible. Here we use a noninteracting degenerate Fermi gas for a proof of principle study, in which
other cooling mechanisms are minimized and the excited high-energy atoms will leave the trap quickly without colliding with the low-energy atoms, manifesting the parametric cooling effect.

Anharmonicity of the trapping potential plays a central role in parametric cooling. In our experiment, we load $^6$Li atoms into a crossed-beam ODT with a crossing angle $2\theta=12^{\circ}$ in the x-z plane, which is described by $U(x,y,z) = -U_0[(1+Z_-^2/z_0^2)^{-1}\,\text{Exp}[-2(y^2+X_+^2)/w_0^2]+(1+Z_+^2/z_0^2)^{-1}\,\text{Exp}[-2(y^2+X_-^2)/w_0^2]]$, with $X_{\pm}=x\cos \theta \pm z\sin \theta$ and $Z_{\pm}=z\cos \theta \pm x\sin \theta$. The single-beam trap depth $U_0$ is 2.8 $\mu$K which is formed by a 100 mW Gaussian beam at 1.06 $\mu$m wavelength with a focused beam waist $w_0=37\,\mu$m. The local radial trap frequency along the x-axis $\omega_x(x,y,z)$ can be calculated from anharmonic radial motions of the atoms~\cite{Poli02}, given by
\begin{equation}
\omega_x(x,y,z)=\frac{\pi\sqrt{2/m}}{\displaystyle \int_{-x}^{x} [U(x,y,z)-U(\widetilde{x},y,z)]^{-1/2}d\widetilde{x}}.
\label{eq:cooling}
\end{equation}
From Eq.\ref{eq:cooling}, the harmonic frequency $\omega_{x0,y0,z0}$ is readily obtained by approximating the center of the trap with a harmonic potential of $m(\omega_{x0}^2 x^2+\omega_{y0}^2 y^2+\omega_{z0}^2 z^2)/2$, where $m$ is the mass of $^6$Li atom. We plot the dispersion of the local frequency $\omega_x(x,0,z)$ in the $y=0$ plane in Fig.~\ref{fig:2DAnH}, showing that the local frequency decreases significantly from the center to the edge of the atom cloud even at $T/T_F=0.6$. The large frequency drops along the z-axis allows applying parametric excitation to selectively remove high-energy atoms in a degenerate Fermi gas.

In the noninteracting regime, parametric modulations usually induce temperature (energy) anisotropy since the atomic motions along the different axes of the clouds are uncoupled. Instead of $T/T_F$, we use $E/E_F$, the ratio between the energy per particle and the Fermi energy, as an effective thermometry, which provides a convenient way to characterize temperature (energy) anisotropy due to the parametric modulation. The total energy per particle is given by $E=E_x+E_y+E_z$ based on uncoupled atomic motions in different directions. For a noninteracting gas, the viral theorem gives $E_{x,y,z}=2\,U_{x,y,z}$ by using an harmonic approximation for the trapping potential. $U_x$ is the potential energy per particle along the x-axis, which can be determined by $U_x= N^{1/3} m \omega_x^2\langle x^2\rangle/2$.  The number-independent mean square size (NIMS) $\langle x^2\rangle=\int x^2 n(x) dx /N^{1/3}$ can be obtained directly from the 1D density profile of the atom cloud~\cite{Luo09T}. Finally the energy in the x-direction is given by $E_x/E_F= m \omega_x^2 \langle x^2\rangle/(6^{1/3}\hbar \overline{\omega})$, where $\overline{\omega}=(\omega_x \omega_y \omega_z)^{1/3}$ is the average trap frequency. It is noted that we ignore the anharmonic correction of the potential energy since it is small for a degenerate Fermi gas at low temperatures.

\begin{figure}[tb]
\includegraphics[width=3.3in]{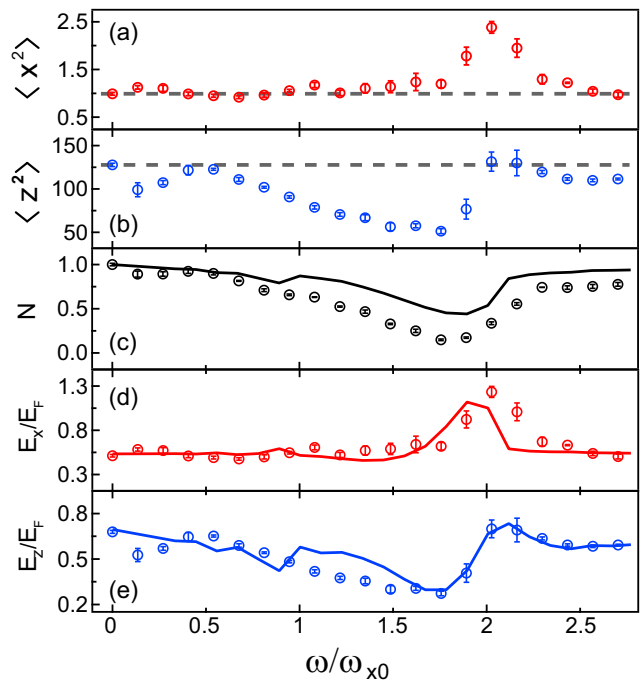}
\caption{The dependence of parametric excitation on the modulation frequency. The radial and axial NIMS, the normalized atom number, and the radial and axial energies are shown from the top to the bottom. All the data have a modulation time $t_m$ = 500 ms and a modulation amplitude $\delta=0.15$. The harmonic frequency $\omega_{x0}=740$ Hz is the measured value. The dashed lines indicate the average value without parametric modulation ($\delta$=0). The solid lines show the simulation results.}
\label{fig:EoverEFvsFreq}
\end{figure}

We prepare a gas of $^6$Li atoms in the two lowest hyperfine states of $F=1/2,m_F=\pm1/2$ (as $|1\rangle$ and $|2\rangle$ states) in a magneto-optical trap. The precooled atoms are then transferred into a crossed-beam ODT made by a 100 W IPG fiber laser. The bias magnetic field is quickly swept to 330 G to implement evaporative cooling. The trap potential is lowered to 0.1$\%$ of the full trap depth in 2.6 s, giving a final trap depth of $5.6\,\mu$K for the crossed-beam trap. A noisy radio-frequency pulse is then applied to prepare a 50:50 spin mixture. To prepare a noninteracting Fermi gas, the magnetic field is swept to 527.3 G, where the s-wave scattering length of $|1\rangle$ and $|2\rangle$ states is zero~\cite{zurn13}. Typically we have a noninteracting Fermi gas of $N=1.6\times10^5$ atoms per spin state at $T/T_F\approx0.6$ with $T_F\approx1.6\,\mu$K to start parametric modulation. The temperature of a noninteracting Fermi gas is measured by fitting the 1D density profile with the finite temperature Thomas-Fermi distribution~\cite{luothesis}. In the noninteracting regime, the temperature from the density profile is also confirmed by fitting the time-of-flight cloud sizes with ballistic expansion dynamics~\cite{tempmesa}.

The parametric excitation is applied to the atom clouds by modulating the optical intensity of the trapping beams with an acousto-optic modulator. The trap depth is modulated as $U_0(t)= U_0[1+\delta \cos (\omega_m t+\theta)]$, where $U_0$ is the trap depth without modulation, $\delta$ is the modulation depth, $\omega_m$ is the modulation frequency, and $\theta$ is the modulation phase. The modulation is turned on for a time $t_m$. After that, the atoms are allowed to stay in the trap for 100 ms to reach a steady state before the trap is turned off. Subsequently the atom cloud ballistically expands for 800 $\mu$s before a 10 $\mu$s resonant optical pulse is applied for absorption imaging. We first extract both the radial and axial NIMSs of the time-of-flight clouds from the absorption images, and then determine the NIMSs and energies of the in-situ clouds from ballistic expansion. We avoid taking in-situ images to eliminate the systematic error due to the high column density.

We first use small $\delta=0.05$ to measure the harmonic trap frequencies, where modulations of twice the harmonic trap frequency induces parametric heating~\cite{savard97}. We measure $\omega_{x0} = 2 \pi\times (740\pm 10)$ Hz, $\omega_{y0} =  2 \pi\times (750\pm 10)$ Hz, $\omega_{z0} = 2 \pi\times (75\pm 5)$ Hz, which agree very well with the theoretical calculation based on the parameters of the trapping potential. We next examine the dependence of the parametric excitation effect on the modulation frequency using large modulation amplitude $\delta=0.15$. The radial (x-axis) and axial (z-axis) NIMSs and energies of the clouds are shown in Fig.~\ref{fig:EoverEFvsFreq}, where $\omega_m$ varies from a near zero value to about $2.5\omega_{x0}$. We find that the NIMSs of the axial and radial directions show quite different frequency dependence, resulting in the energy anisotropy after modulation. The radial NIMSs increase significantly around $2\omega_{x0}$, indicating the usual parametric heating effect~\cite{savard97,Friebel98}. In contrast, the axial NIMSs barely change at $2\omega_{x0}$, showing the modulation at $2\omega_{x0}$ frequency mainly excites atomic motion along the radial direction and does not couple to the motion along the axial direction. The most striking feature evident in our measurements is that the axial NIMSs decrease significantly in a wide range between $0.6\,\omega_{x0}$ to $1.7\,\omega_{x0}$, indicating a reduction of the axial cloud energy $E_z$. This axial parametric cooling effect can be explained by the fact that the atoms along the z-axis experience different local radial frequency $\omega_{x}(z)$, such that the high-energy atoms at the edge of the trap are selectively excited out of the trap. As shown in Fig.~\ref{fig:2DAnH}, anharmonicity plays a much more significant role in the axial direction than that in the radial one, therefore parametric cooling mainly takes place in the axial direction.

We develop a simple model of a group of anharmonic oscillators distributed in a 2D plane to simulate the parametric modulation process. We assume the initial column density is described by a 2D Thomas-Fermi $n(x,z)$ distribution~\cite{luothesis}, and the local densities $n(x,z)$ are associated to different anharmonic oscillators. During the parametric modulation, the atoms associated to a specific $n(x,z)$ oscillate along the radial (x-axis) direction only, whose equation of motion is given by
\begin{equation}
\label{eq:Motion}
\frac{d^2x}{dt^2}+ \frac{1+\delta\cos(\omega_m t+\theta)}{m}\frac{dU(x,z)}{dx}=0\,\,.
\end{equation}
Here $U(x,z)=U(x,0,z)$ by approximating the column density in the 2D plane of $y=0$. The initial atom position and velocity are given by $x(0)=x_0$ and $(dx/dt)_{t=0}=(2U(x_0,0,z)-2U(0,0,z))/2m)^{1/2}$. We solve Eq.~\ref{eq:Motion} numerically for $t \in \{0,t_m\}$  to obtain the position and kinetic energy of atoms after parametric excitation. In this model, atoms associated with a local density $n(x,z)$ can be selectively excited when the parametric oscillation frequency $\omega_m$ is tuned to resonance with the local frequency $\omega_x(x,z)$. When $\omega_m\,\approx\,2\,\omega_{x0}$, the modulation will excite the atoms in the center of the trap, inducing a loss of the low-energy atoms and an increase of the energy per particle. In comparison, when the $\omega_m$ is tuned to $1.5\,\omega_{x0}$ close to the local trapping frequency at the edge of the trap, the parametric process will excite the high-energy atoms out of the trap and result in a cooling effect. The simulation results of the atom number, $E_x$, and $E_z$ are shown by the solid lines in Fig.~\ref{fig:EoverEFvsFreq}, where we only adjust the harmonic frequency $\omega_{x0}$ to 825 Hz in our simulation for the best fitting of the experimental data, while keeping all other simulation parameters as the experimental values. This model manifests the analogy between parametric excitation in an optical trap and rf-knife in a magnetic trap, both of which depend on the spatial variation of frequency (either trapping frequency or rf transition frequency) between high-energy and low-energy atoms.

\begin{figure}[tb]
\includegraphics[width=3.5in]{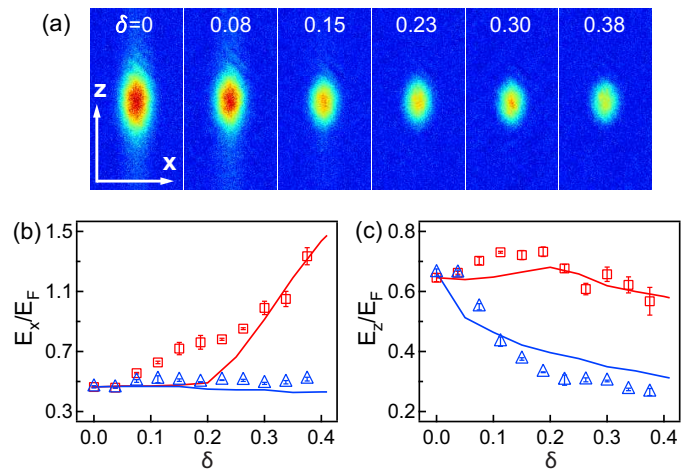}
\caption{The dependence of the radial and axial energies on the modulation amplitudes.  (a) The absorption images of the atoms clouds show a dramatic decrease of the axial cloud sizes with an increase of modulation amplitudes, where $\omega_m=1.5\,\omega_{x0}$. (b) The dependence of the radial energies on the modulation amplitude. (c) The dependence of the radial energies on the modulation amplitude. In both (b) and (c), blue triangles are at the modulation frequency $1.5\omega_{x0}$, and red squares are at $2.0\omega_{x0}$. The solid lines represent the simulation of the anharmonic oscillator model using the same simulation parameters for the frequency dependence.}
\label{fig:EoverEFvsAmp}
\end{figure}

We also study the dependence of the parametric excitation on the modulation amplitude with the results shown in Fig.~\ref{fig:EoverEFvsAmp}. For the modulation $\omega_{m}=2\,\omega_{x0}= 2\pi \times 1500$ Hz, $E_x/E_F$ increases dramatically when the modulation amplitude increases, which is consistent with the parametric heating effect along the radial direction. For the modulation  $\omega_{m}=1.5\,\omega_{x0}$, $E_z/E_F$ decreases significantly with an increase of the modulation amplitude, showing that a stronger cooling effect takes place when larger modulation expels more high-energy atoms out of the trap. The cooling effect saturates when $\delta$ increases to 0.25 due to the fact that the modulation becomes so strong that most atoms in the anharmonic region have already been expelled from the trap. We simulate the dependence on the modulation amplitude shown by the solid lines in Fig.\ref{fig:EoverEFvsAmp} (b) and (c). The simulations exhibit both heating and cooling features, which agree with the experimental results reasonably well.

After parametric modulation, the anisotropic energy distributions between the axial and radial directions are observed. We find that these anisotropic energy distributions are quasi-steady around 528 G, showing a lack of cross-dimensional thermalization~\cite{Monroe98,Aikawa14,hahnthesis} in the noninteracting regime. For a noninteracting Fermi gas, s-wave elastic collisions are absent, and the cross-dimensional term of the trapping potential is negligible for a degenerate Fermi gas at very low temperatures. To obtain isotropic temperature, the magnetic field is adiabatically swept to 330 G, where a finite scattering length of -280 $a_0$~\cite{Duarte11} assists a fast cross-dimensional thermalization. After that, the atom clouds are swept back to 527.3 G, and the energy distributions are measured by taking absorption images. We verify that the equipartition energy distribution is retrieved within a 100 ms holding time at 330 G. By subtracting the finite heating rate during the sweeping and holding time~\cite{TrapHL}, we find that the energy per particle $E$ of a noninteracting Fermi gas is decreased from $1.6\,E_F$ to $1.3\,E_F$ by parametric cooling. It is worth noting that this parametric cooling effect is almost optimal since the cloud energy decreases only in the axial direction. For a noninteracting Fermi gas in a harmonic trap, the ground state energy in each direction is $0.25\,E_F$. Starting with the initial cloud energy $E=1.60\,E_F$, the coldest sample that can be obtained by axial parametric cooling is limited by $(2/3)1.60E_F+0.25E_F=1.3E_F$.

In conclusion, we report parametric cooling of a noninteracting degenerate Fermi gas. The parametric cooling method provides a selective way to remove high-energy atoms from optical traps in the noninteracting regime. It also provides a convenient method to generate temperature anisotropy for studying nonequilibrium thermodynamics in quantum gases. This method can be implemented in many existing optical traps, such as a power-law trap~\cite{Bruce11} and a box trap~\cite{Gaunt13}, in which large anharmonicity exists in all three directions. With specially designed anharmonic traps, parametric cooling has potential to be implemented in a large temperature scale, where a Fermi gas may be directly cooled from the thermal state to the degenerate regime by parametric modulations.

Le Luo is a member of the Indiana University Center for Spacetime Symmetries (IUCSS). Le Luo thanks supports from Indiana University RSFG and Purdue University PRF.


\begin{thebibliography}{26}
\expandafter\ifx\csname natexlab\endcsname\relax\def\natexlab#1{#1}\fi
\expandafter\ifx\csname bibnamefont\endcsname\relax
  \def\bibnamefont#1{#1}\fi
\expandafter\ifx\csname bibfnamefont\endcsname\relax
  \def\bibfnamefont#1{#1}\fi
\expandafter\ifx\csname citenamefont\endcsname\relax
  \def\citenamefont#1{#1}\fi
\expandafter\ifx\csname url\endcsname\relax
  \def\url#1{\texttt{#1}}\fi
\expandafter\ifx\csname urlprefix\endcsname\relax\def\urlprefix{URL }\fi
\providecommand{\bibinfo}[2]{#2}
\providecommand{\eprint}[2][]{\url{#2}}

\bibitem[{\citenamefont{Barrett et~al.}(2001)\citenamefont{Barrett, Sauer, and
  Chapman}}]{ChampmanBEC}
\bibinfo{author}{\bibfnamefont{M.~D.} \bibnamefont{Barrett}},
  \bibinfo{author}{\bibfnamefont{J.~A.} \bibnamefont{Sauer}}, \bibnamefont{and}
  \bibinfo{author}{\bibfnamefont{M.~S.} \bibnamefont{Chapman}},
  \bibinfo{journal}{Phys. Rev. Lett.} \textbf{\bibinfo{volume}{87}},
  \bibinfo{pages}{010404} (\bibinfo{year}{2001}).

\bibitem[{\citenamefont{Granade et~al.}(2002)\citenamefont{Granade, Gehm,
  O'Hara, and Thomas}}]{GranadeAllopt}
\bibinfo{author}{\bibfnamefont{S.~R.} \bibnamefont{Granade}},
  \bibinfo{author}{\bibfnamefont{M.~E.} \bibnamefont{Gehm}},
  \bibinfo{author}{\bibfnamefont{K.~M.} \bibnamefont{O'Hara}},
  \bibnamefont{and} \bibinfo{author}{\bibfnamefont{J.~E.}
  \bibnamefont{Thomas}}, \bibinfo{journal}{Phys. Rev. Lett.}
  \textbf{\bibinfo{volume}{88}}, \bibinfo{pages}{120405}
  (\bibinfo{year}{2002}).

\bibitem[{\citenamefont{Grimm. et~al.}(2000)\citenamefont{Grimm.,
  Weidem\mbox{\"{u}}ller, and Ovchinnikov}}]{Grimmreview}
\bibinfo{author}{\bibfnamefont{R.}~\bibnamefont{Grimm.}},
  \bibinfo{author}{\bibfnamefont{M.}~\bibnamefont{Weidem\mbox{\"{u}}ller}},
  \bibnamefont{and} \bibinfo{author}{\bibfnamefont{Y.~B.}
  \bibnamefont{Ovchinnikov}}, \bibinfo{journal}{Adv. At. Mol. Opt. Phys.}
  \textbf{\bibinfo{volume}{42}}, \bibinfo{pages}{95} (\bibinfo{year}{2000}).

\bibitem[{\citenamefont{Cl\'ement et~al.}(2009)\citenamefont{Cl\'ement,
  Brantut, Robert-de Saint-Vincent, Nyman, Aspect, Bourdel, and
  Bouyer}}]{Clement09}
\bibinfo{author}{\bibfnamefont{J.-F.} \bibnamefont{Cl\'ement}},
  \bibinfo{author}{\bibfnamefont{J.-P.} \bibnamefont{Brantut}},
  \bibinfo{author}{\bibfnamefont{M.}~\bibnamefont{Robert-de Saint-Vincent}},
  \bibinfo{author}{\bibfnamefont{R.~A.} \bibnamefont{Nyman}},
  \bibinfo{author}{\bibfnamefont{A.}~\bibnamefont{Aspect}},
  \bibinfo{author}{\bibfnamefont{T.}~\bibnamefont{Bourdel}}, \bibnamefont{and}
  \bibinfo{author}{\bibfnamefont{P.}~\bibnamefont{Bouyer}},
  \bibinfo{journal}{Phys. Rev. A} \textbf{\bibinfo{volume}{79}},
  \bibinfo{pages}{061406} (\bibinfo{year}{2009}).

\bibitem[{\citenamefont{Kinoshita et~al.}(2005)\citenamefont{Kinoshita, Wenger,
  and Weiss}}]{Kinoshita05}
\bibinfo{author}{\bibfnamefont{T.}~\bibnamefont{Kinoshita}},
  \bibinfo{author}{\bibfnamefont{T.}~\bibnamefont{Wenger}}, \bibnamefont{and}
  \bibinfo{author}{\bibfnamefont{D.~S.} \bibnamefont{Weiss}},
  \bibinfo{journal}{Phys. Rev. A} \textbf{\bibinfo{volume}{71}},
  \bibinfo{pages}{011602} (\bibinfo{year}{2005}).

\bibitem[{\citenamefont{Arnold and Barrett}(2011)}]{Arnold11}
\bibinfo{author}{\bibfnamefont{K.}~\bibnamefont{Arnold}} \bibnamefont{and}
  \bibinfo{author}{\bibfnamefont{M.}~\bibnamefont{Barrett}},
  \bibinfo{journal}{Optics Communications} \textbf{\bibinfo{volume}{284}},
  \bibinfo{pages}{3288 } (\bibinfo{year}{2011}).

\bibitem[{\citenamefont{Hung et~al.}(2008)\citenamefont{Hung, Zhang, Gemelke,
  and Chin}}]{Hung08}
\bibinfo{author}{\bibfnamefont{C.-L.} \bibnamefont{Hung}},
  \bibinfo{author}{\bibfnamefont{X.}~\bibnamefont{Zhang}},
  \bibinfo{author}{\bibfnamefont{N.}~\bibnamefont{Gemelke}}, \bibnamefont{and}
  \bibinfo{author}{\bibfnamefont{C.}~\bibnamefont{Chin}},
  \bibinfo{journal}{Phys. Rev. A} \textbf{\bibinfo{volume}{78}},
  \bibinfo{pages}{011604} (\bibinfo{year}{2008}).

\bibitem[{\citenamefont{Ketterle and Druten}(1996)}]{KetterleVanDruten}
\bibinfo{author}{\bibfnamefont{W.}~\bibnamefont{Ketterle}} \bibnamefont{and}
  \bibinfo{author}{\bibfnamefont{N.~J.~V.} \bibnamefont{Druten}},
  \bibinfo{journal}{Adv. At. Mol. Opt. Phys.} \textbf{\bibinfo{volume}{37}},
  \bibinfo{pages}{181} (\bibinfo{year}{1996}).

\bibitem[{\citenamefont{Luo et~al.}(2006)\citenamefont{Luo, Clancy, Joseph,
  Kinast, A.Turlapov, and Thomas}}]{Luo06Cooling}
\bibinfo{author}{\bibfnamefont{L.}~\bibnamefont{Luo}},
  \bibinfo{author}{\bibfnamefont{B.}~\bibnamefont{Clancy}},
  \bibinfo{author}{\bibfnamefont{J.}~\bibnamefont{Joseph}},
  \bibinfo{author}{\bibfnamefont{J.}~\bibnamefont{Kinast}},
  \bibinfo{author}{\bibnamefont{A.Turlapov}}, \bibnamefont{and}
  \bibinfo{author}{\bibfnamefont{J.~E.} \bibnamefont{Thomas}},
  \bibinfo{journal}{New Journal of Physics} \textbf{\bibinfo{volume}{8}},
  \bibinfo{pages}{213} (\bibinfo{year}{2006}).

\bibitem[{\citenamefont{Moses et~al.}(2015)\citenamefont{Moses, Covey,
  Miecnikowski, Yan, Gadway, Ye, and Jin}}]{Moses15}
\bibinfo{author}{\bibfnamefont{S.~A.} \bibnamefont{Moses}},
  \bibinfo{author}{\bibfnamefont{J.~P.} \bibnamefont{Covey}},
  \bibinfo{author}{\bibfnamefont{M.~T.} \bibnamefont{Miecnikowski}},
  \bibinfo{author}{\bibfnamefont{B.}~\bibnamefont{Yan}},
  \bibinfo{author}{\bibfnamefont{B.}~\bibnamefont{Gadway}},
  \bibinfo{author}{\bibfnamefont{J.}~\bibnamefont{Ye}}, \bibnamefont{and}
  \bibinfo{author}{\bibfnamefont{D.~S.} \bibnamefont{Jin}},
  \bibinfo{journal}{Science} \textbf{\bibinfo{volume}{350}},
  \bibinfo{pages}{659} (\bibinfo{year}{2015}).

\bibitem[{\citenamefont{Luo}(2008)}]{luothesis}
\bibinfo{author}{\bibfnamefont{L.}~\bibnamefont{Luo}}, Ph.D. thesis,
  \bibinfo{school}{Duke University} (\bibinfo{year}{2008}).

\bibitem[{\citenamefont{Savard et~al.}(1997)\citenamefont{Savard, O'Hara, and
  Thomas}}]{savard97}
\bibinfo{author}{\bibfnamefont{T.~A.} \bibnamefont{Savard}},
  \bibinfo{author}{\bibfnamefont{K.~M.} \bibnamefont{O'Hara}},
  \bibnamefont{and} \bibinfo{author}{\bibfnamefont{J.~E.}
  \bibnamefont{Thomas}}, \bibinfo{journal}{Phys. Rev. A}
  \textbf{\bibinfo{volume}{56}}, \bibinfo{pages}{R1095} (\bibinfo{year}{1997}).

\bibitem[{\citenamefont{Gehm et~al.}(1998)\citenamefont{Gehm, O'Hara, Savard,
  and Thomas}}]{gehm98}
\bibinfo{author}{\bibfnamefont{M.~E.} \bibnamefont{Gehm}},
  \bibinfo{author}{\bibfnamefont{K.~M.} \bibnamefont{O'Hara}},
  \bibinfo{author}{\bibfnamefont{T.~A.} \bibnamefont{Savard}},
  \bibnamefont{and} \bibinfo{author}{\bibfnamefont{J.~E.}
  \bibnamefont{Thomas}}, \bibinfo{journal}{Phys. Rev. A}
  \textbf{\bibinfo{volume}{58}}, \bibinfo{pages}{3914} (\bibinfo{year}{1998}).

\bibitem[{\citenamefont{Kumakura et~al.}(2003)\citenamefont{Kumakura,
  Shirahata, Takasu, Takahashi, and Yabuzaki}}]{Kumakura03}
\bibinfo{author}{\bibfnamefont{M.}~\bibnamefont{Kumakura}},
  \bibinfo{author}{\bibfnamefont{Y.}~\bibnamefont{Shirahata}},
  \bibinfo{author}{\bibfnamefont{Y.}~\bibnamefont{Takasu}},
  \bibinfo{author}{\bibfnamefont{Y.}~\bibnamefont{Takahashi}},
  \bibnamefont{and} \bibinfo{author}{\bibfnamefont{T.}~\bibnamefont{Yabuzaki}},
  \bibinfo{journal}{Phys. Rev. A} \textbf{\bibinfo{volume}{68}},
  \bibinfo{pages}{021401(R)} (\bibinfo{year}{2003}).

\bibitem[{\citenamefont{Poli et~al.}(2002)\citenamefont{Poli, Brecha, Roati,
  and Modugno}}]{Poli02}
\bibinfo{author}{\bibfnamefont{N.}~\bibnamefont{Poli}},
  \bibinfo{author}{\bibfnamefont{R.~J.} \bibnamefont{Brecha}},
  \bibinfo{author}{\bibfnamefont{G.}~\bibnamefont{Roati}}, \bibnamefont{and}
  \bibinfo{author}{\bibfnamefont{G.}~\bibnamefont{Modugno}},
  \bibinfo{journal}{Phys. Rev. A} \textbf{\bibinfo{volume}{65}},
  \bibinfo{pages}{021401} (\bibinfo{year}{2002}).

\bibitem[{\citenamefont{Luo and Thomas}(2008)}]{Luo09T}
\bibinfo{author}{\bibfnamefont{L.}~\bibnamefont{Luo}} \bibnamefont{and}
  \bibinfo{author}{\bibfnamefont{J.~E.} \bibnamefont{Thomas}},
  \bibinfo{journal}{Journal of Low Temperature Physics}
  \textbf{\bibinfo{volume}{154}}, \bibinfo{pages}{1} (\bibinfo{year}{2008}).

\bibitem[{\citenamefont{Z\"urn et~al.}(2013)\citenamefont{Z\"urn, Lompe, Wenz,
  Jochim, Julienne, and Hutson}}]{zurn13}
\bibinfo{author}{\bibfnamefont{G.}~\bibnamefont{Z\"urn}},
  \bibinfo{author}{\bibfnamefont{T.}~\bibnamefont{Lompe}},
  \bibinfo{author}{\bibfnamefont{A.~N.} \bibnamefont{Wenz}},
  \bibinfo{author}{\bibfnamefont{S.}~\bibnamefont{Jochim}},
  \bibinfo{author}{\bibfnamefont{P.~S.} \bibnamefont{Julienne}},
  \bibnamefont{and} \bibinfo{author}{\bibfnamefont{J.~M.}
  \bibnamefont{Hutson}}, \bibinfo{journal}{Phys. Rev. Lett.}
  \textbf{\bibinfo{volume}{110}}, \bibinfo{pages}{135301}
  (\bibinfo{year}{2013}).

\bibitem[{tem()}]{tempmesa}
\bibinfo{note}{The ballistic expansion of a noninteracting Fermi gas provides
  $\langle x^2(t)\rangle=\langle x^2_0\rangle[1+2k_BTt^2/(m\langle
  x^2_0\rangle)]$, where $t$ is time-of flight and $\langle x^2(t)\rangle$ is
  the mean square cloud size. The temperature $T$ can be determined by the time
  dependence of the mean-square size.}

\bibitem[{\citenamefont{Friebel et~al.}(1998)\citenamefont{Friebel, D'Andrea,
  Walz, Weitz, and H\mbox{\"{a}}nsch}}]{Friebel98}
\bibinfo{author}{\bibfnamefont{S.}~\bibnamefont{Friebel}},
  \bibinfo{author}{\bibfnamefont{C.}~\bibnamefont{D'Andrea}},
  \bibinfo{author}{\bibfnamefont{J.}~\bibnamefont{Walz}},
  \bibinfo{author}{\bibfnamefont{M.}~\bibnamefont{Weitz}}, \bibnamefont{and}
  \bibinfo{author}{\bibfnamefont{T.~W.} \bibnamefont{H\mbox{\"{a}}nsch}},
  \bibinfo{journal}{Phys. Rev. A} \textbf{\bibinfo{volume}{57}},
  \bibinfo{pages}{R20} (\bibinfo{year}{1998}).

\bibitem[{\citenamefont{Monroe et~al.}(1993)\citenamefont{Monroe, Cornell,
  Sackett, Myatt, and Wieman}}]{Monroe98}
\bibinfo{author}{\bibfnamefont{C.~R.} \bibnamefont{Monroe}},
  \bibinfo{author}{\bibfnamefont{E.~A.} \bibnamefont{Cornell}},
  \bibinfo{author}{\bibfnamefont{C.~A.} \bibnamefont{Sackett}},
  \bibinfo{author}{\bibfnamefont{C.~J.} \bibnamefont{Myatt}}, \bibnamefont{and}
  \bibinfo{author}{\bibfnamefont{C.~E.} \bibnamefont{Wieman}},
  \bibinfo{journal}{Phys. Rev. Lett.} \textbf{\bibinfo{volume}{70}},
  \bibinfo{pages}{414} (\bibinfo{year}{1993}).

\bibitem[{\citenamefont{Aikawa et~al.}(2014)\citenamefont{Aikawa, Frisch, Mark,
  Baier, Grimm, Bohn, Jin, Bruun, and Ferlaino}}]{Aikawa14}
\bibinfo{author}{\bibfnamefont{K.}~\bibnamefont{Aikawa}},
  \bibinfo{author}{\bibfnamefont{A.}~\bibnamefont{Frisch}},
  \bibinfo{author}{\bibfnamefont{M.}~\bibnamefont{Mark}},
  \bibinfo{author}{\bibfnamefont{S.}~\bibnamefont{Baier}},
  \bibinfo{author}{\bibfnamefont{R.}~\bibnamefont{Grimm}},
  \bibinfo{author}{\bibfnamefont{J.}~\bibnamefont{Bohn}},
  \bibinfo{author}{\bibfnamefont{D.}~\bibnamefont{Jin}},
  \bibinfo{author}{\bibfnamefont{G.}~\bibnamefont{Bruun}}, \bibnamefont{and}
  \bibinfo{author}{\bibfnamefont{F.}~\bibnamefont{Ferlaino}},
  \bibinfo{journal}{Phys. Rev. Lett.} \textbf{\bibinfo{volume}{113}},
  \bibinfo{pages}{263201} (\bibinfo{year}{2014}).

\bibitem[{\citenamefont{Hahn}(2009)}]{hahnthesis}
\bibinfo{author}{\bibfnamefont{C.}~\bibnamefont{Hahn}}, Master's thesis,
  \bibinfo{school}{Ludwig Maximilians University Munich}
  (\bibinfo{year}{2009}).

\bibitem[{\citenamefont{Duarte et~al.}(2011)\citenamefont{Duarte, Hart,
  Hitchcock, Corcovilos, Yang, Reed, and Hulet}}]{Duarte11}
\bibinfo{author}{\bibfnamefont{P.~M.} \bibnamefont{Duarte}},
  \bibinfo{author}{\bibfnamefont{R.~A.} \bibnamefont{Hart}},
  \bibinfo{author}{\bibfnamefont{J.~M.} \bibnamefont{Hitchcock}},
  \bibinfo{author}{\bibfnamefont{T.~A.} \bibnamefont{Corcovilos}},
  \bibinfo{author}{\bibfnamefont{T.-L.} \bibnamefont{Yang}},
  \bibinfo{author}{\bibfnamefont{A.}~\bibnamefont{Reed}}, \bibnamefont{and}
  \bibinfo{author}{\bibfnamefont{R.~G.} \bibnamefont{Hulet}},
  \bibinfo{journal}{Phys. Rev. Lett.} \textbf{\bibinfo{volume}{84}},
  \bibinfo{pages}{061406(R)} (\bibinfo{year}{2011}).

\bibitem[{Tra()}]{TrapHL}
\bibinfo{note}{The trap lifetime is around 20 seconds, where the heating and
  loss is due to the background collisions and/or trapping potential noise. To
  manifest the rethermalization, we subtract this heating effect due to the
  finite magnetic field sweeping and holding time.}

\bibitem[{\citenamefont{Bruce et~al.}(2011)\citenamefont{Bruce, Bromley,
  Smirne, Torralbo-Campo, and Cassettari}}]{Bruce11}
\bibinfo{author}{\bibfnamefont{G.~D.} \bibnamefont{Bruce}},
  \bibinfo{author}{\bibfnamefont{S.~L.} \bibnamefont{Bromley}},
  \bibinfo{author}{\bibfnamefont{G.}~\bibnamefont{Smirne}},
  \bibinfo{author}{\bibfnamefont{L.}~\bibnamefont{Torralbo-Campo}},
  \bibnamefont{and}
  \bibinfo{author}{\bibfnamefont{D.}~\bibnamefont{Cassettari}},
  \bibinfo{journal}{Phys. Rev. A} \textbf{\bibinfo{volume}{84}},
  \bibinfo{pages}{053410} (\bibinfo{year}{2011}).

\bibitem[{\citenamefont{Gaunt et~al.}(2013)\citenamefont{Gaunt, Schmidutz,
  Gotlibovych, Smith, and Hadzibabic}}]{Gaunt13}
\bibinfo{author}{\bibfnamefont{A.~L.} \bibnamefont{Gaunt}},
  \bibinfo{author}{\bibfnamefont{T.~F.} \bibnamefont{Schmidutz}},
  \bibinfo{author}{\bibfnamefont{I.}~\bibnamefont{Gotlibovych}},
  \bibinfo{author}{\bibfnamefont{R.~P.} \bibnamefont{Smith}}, \bibnamefont{and}
  \bibinfo{author}{\bibfnamefont{Z.}~\bibnamefont{Hadzibabic}},
  \bibinfo{journal}{Phys. Rev. Lett.} \textbf{\bibinfo{volume}{110}},
  \bibinfo{pages}{200406} (\bibinfo{year}{2013}).

\end{thebibliography}

\end{document}